\documentclass[letterpaper, 10 pt, conference]{ieeeconf}  
\usepackage{cite}
\usepackage{amssymb,amsfonts,amsmath}
\usepackage{array}
\usepackage{bm}
\usepackage{textcomp} 
\usepackage{xcolor}
\usepackage{csquotes}
\usepackage{color}
\usepackage{comment}
\usepackage{mdwmath}
\usepackage{centernot}
\usepackage{mdwtab}
\usepackage{ifpdf}
\usepackage{eqparbox}
\usepackage{bbold}
\usepackage[font=footnotesize]{subfig}
\usepackage{fixltx2e}
\usepackage{stfloats}
   
\DeclareMathOperator*{\argmax}{argmax}
\usepackage{hyperref}
\usepackage[pdftex]{graphicx}
\usepackage[linesnumbered,commentsnumbered,ruled,vlined]{algorithm2e}
\usepackage{url}
\usepackage{float}
\usepackage{lipsum}
\usepackage{algpseudocode}
\usepackage{listings}
\usepackage{multirow}
\newtheorem{assumption}{\rm\textbf{Assumption}}

\newtheorem{definition}{\rm\textbf{Definition}}
\newtheorem{theorem}{\rm\textbf{Theorem}}
\newtheorem{lemma}{\rm\textbf{Lemma}}

\newtheorem{remark}{\rm\textbf{Remark}}

\IEEEoverridecommandlockouts                              

\title{\LARGE \bf
Trust-Aware Resilient Control and Coordination\\ of Connected and Automated Vehicles 
}

\author{H.M. Sabbir Ahmad$^{1}$, Ehsan Sabouni$^{1}$, Wei Xiao$^2$, Christos G. Cassandras$^1$ and Wenchao Li$^1$
\thanks{$^1$Division of Systems Engineering and Department of Electrical \& Computer Engineering, Boston University, Boston, MA, USA, \texttt{{\small \{sabbir92, esabouni, cgc, wenchao\}@bu.edu}}}
\thanks{$^2$Computer Science \& Artificial Intelligence Laboratory, Massachusetts Institute of Technology, Cambridge, MA, USA \texttt{{\small \{weixy@mit.edu\} }}}
\thanks{This work was supported in part by NSF under grants ECCS-1931600,
DMS-1664644, CNS-1645681, CNS-2149511, by AFOSR under grant FA9550-19-1-0158,
by ARPA-E under grant DE-AR0001282, by the MathWorks and by NPRP grant
(12S-0228-190177) from the Qatar National Research Fund, a member of
the Qatar Foundation (the statements made herein are solely the responsibility
of the authors).}
}%

\newif\ifarvix
\arvixfalse

\newif\ifread  
\readfalse

\newif\ifITSC
\ITSCtrue

\begin{document}
\lstset{language=C}
\maketitle
\thispagestyle{empty}
\pagestyle{empty}

\begin{abstract}
We address the security of a network of Connected and Automated Vehicles (CAVs) cooperating to navigate through a conflict area.  Adversarial attacks such as Sybil attacks can cause safety violations resulting in collisions and traffic jams.  In addition, uncooperative (but not necessarily adversarial) CAVs can also induce similar adversarial effects on the traffic network.  We propose a decentralized resilient control and coordination scheme that mitigates the effects of adversarial attacks and uncooperative CAVs by utilizing a trust framework.  Our trust-aware scheme can guarantee safe collision free coordination and mitigate traffic jams. Simulation results validate the theoretical guarantee of our proposed scheme, and demonstrate that it can effectively mitigate adversarial effects across different traffic scenarios.
\end{abstract}

\section{INTRODUCTION}
The rise of connected and automated vehicles (CAVs) and advancements in traffic infrastructure \cite{li2013survey} promise to offer solutions to transportation issues like accidents, congestion, energy consumption, and pollution \cite{Schrank20152015UM,kavalchuk2020performance}. To achieve these benefits, efficient traffic management is crucial, particularly at bottleneck locations such as intersections, roundabouts, and merging roadways \cite{VANDENBERG201643}.

Thus far, two approaches, centralized \cite{Liu_01} and decentralized \cite{Xiao_03}, have been proposed for controlling and coordinating CAVs at conflict points. 
There has been extensive research on cybersecurity of CAVs summarized in \cite{Shukla_01,Sun_01,Pham_01}. The attacks can be categorized into in-vehicle network attacks and attacks on (V2V or V2X) communication networks \cite{Sun_01}. A significant amount of research has been done from a control point of view with the aim of designing smart and efficient coordination algorithms for real-world implementation. However, security for this next generation of CAV algorithms has received virtually no attention, with only \cite{Jarouf_01, Zhao_01} tackling security for merging roadways, and our previous work \cite{Ahmad_01} providing an extensive study of security threats to this class of algorithms for various conflict areas. 

There is literature that considers cyberattacks on connected vehicles and investigates their effects on intersections \cite{Huang_01,Chen_01} and freeway \cite{Reilly_01} control systems; however, the fundamental difference is that they do not consider the security of cooperative control of CAVs. One class of cooperative algorithms for autonomous vehicles whose security has been extensively studied \cite{Biroon_02,Boddupalli_01,Farivar_01} is Cooperative Adaptive Cruise Control (CACC).



An idea that has been extensively applied to multi-agent systems is the notion of trust/reputation \cite{Chen_01, Cheng_02, Hu_01}. 
A novel CBF-based trust metric was introduced in \cite{Parwana_01} for multi-robot systems (MRS) for providing safe control against adversarial agents; however, it cannot be directly applied to our application. The authors in \cite{Garlichs_01} used a trust framework to address the security of CACC. Lastly, the authors in \cite{Shoukry_01} used a trust framework based on a macroscopic model of the network to tackle Sybil attacks for traffic intersections without analyzing the fidelity of the model and commenting about the classification accuracy of their proposed method. 

In this paper, we present distributed resilient control and coordination scheme for CAVs at conflict areas that is resilient to adversarial agents and uncooperative CAVs. We use Sybil attacks to validate our proposed scheme as they can be used to achieve both adversarial objectives. Sybil attacks can't be tackled using existing road infrastructure including namely sensors and cameras as they are placed sparsely in the network, and, their reliability degrades with age \cite{Shoukry_01}. 
The key contributions of the paper are as follows. 

\begin{enumerate}
    \item  We propose trust-aware resilient control and coordination that guarantees safe coordination against adversarial attacks and uncooperative CAVs. It is important to add that, our proposed framework is agnostic to the specific implementation of the trust framework. 

    \item We provide resilient coordination using a \emph{robust event-driven scheduling scheme} that can successfully alleviate traffic holdups due to adversarial attacks and uncooperative CAVs. 

    \item We present simulation results that validate our proposed resilient control and coordination scheme guarantees safety; and our robust scheduling scheme besides mitigating traffic jams also improves the travel time and fuel economy of real cooperative CAVs in the presence of adversarial attacks and uncooperative CAVs. 
    
\end{enumerate}

Our proposed scheme is computationally tractable, minimally invasive, and can be readily incorporated into the existing intelligent traffic infrastructure like intersections, roundabouts, merging roadways, etc. without extensive overhaul. 
The paper is organized in seven sections. We present the background materials and the threat models in sections \ref{problem_formulation} and \ref{threat_model} respectively. In section \ref{trust_framework}, we present the trust framework for a cooperative network of CAVs in conflict areas. Our proposed resilient control and coordination scheme is presented in section \ref{resilient_control_formulation}. The results from our simulations have been included in section \ref{Results} which is followed by the conclusion in section \ref{conclusion}.

\section{Background} 
\label{problem_formulation}
\begin{figure*}
\centering
\includegraphics[scale = 0.9]{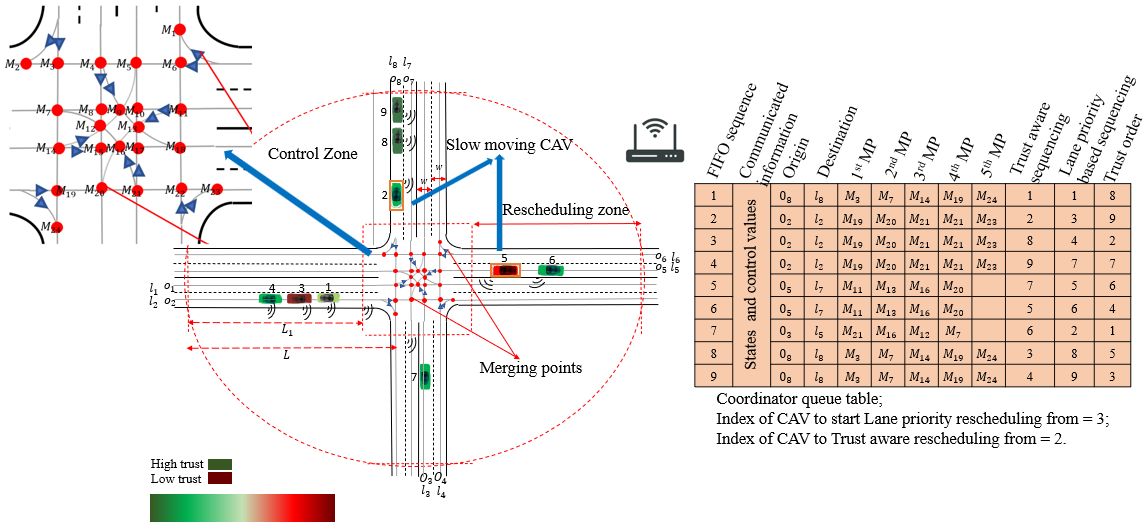}
\caption{The multi-lane intersection problem. Collisions may happen at the merging points. The table shows the order of the CAVs in the queue based on the FIFO sequencing scheme, trust-aware scheduling scheme, and lane-priority based scheduling scheme. 
}
\label{fig:intersection}%
\end{figure*}

We present resilient control and coordination approach for secure coordination of CAVs in conflict areas, \textit{using the signal-free intersection presented in \cite{Xu_02} as an illustrative example}. Fig. \ref{fig:intersection} shows a typical intersection with multiple lanes. The Control Zone (CZ) is the area within the outer red circle. It contains eight entries labeled from $o_1$ to $o_8$ and lanes labeled from $l_1$ to $l_8$ 
each of length $L$ which is assumed to be the same here. Red dots show all the merging points (MPs)
where potential collisions may occur. All the CAVs have the following possible planned trajectories when they enter the CZ: going straight, turning left from the leftmost lane, or turning right from the rightmost lane. 


The vehicle dynamics for each CAV in the CZ take the following form:
\begin{equation}
\left[
\begin{array}
[c]{c}%
\dot{x}_{i}(t)\\
\dot{v}_{i}(t)
\end{array}
\right] =\left[
\begin{array}
[c]{c}%
v_{i}(t)\\
u_{i}(t)
\end{array}
\right], \label{VehicleDynamics}%
\end{equation}
where $x_{i}(t)$ is the distance from the origin at which CAV $i$ arrives, $v_{i}(t)$ and $u_{i}(t)$ denote the velocity and control input (acceleration/deceleration) of CAV $i$, respectively. We also consider that each CAV has a vision-based perception capability defined by a radius and angle tuple denoted as $(r,\theta)$, (where $r \in \mathbb{R}^+, \theta \in [0, 2\pi]$) 
Let $t_{i}^{0}$ and $t_{i}^{f}$ denote the time that CAV $i$ arrives at the origin and exits the CZ, respectively. The control is implemented in a \textit{decentralized manner} whereby each CAV $i$ determines a control policy to jointly minimize the travel time and energy consumption governed by the dynamics (\ref{VehicleDynamics}). Expressing energy through $\frac{1}{2}u_i^2(t)$ and normalizing travel time and energy, we use the weight $\alpha\in[0,1]$ to construct a convex combination as follows:

\begin{equation}\label{eqn:energyobja}
J_{i}(u_{i}(t),t_i^f):=
{\beta(t_{i}^{f}-t_{i}^{0})}
+
{\int_{t_{i}^{0}}^{t_{i}^{f}}\frac{1}{2}u_{i}^{2}(t)dt}
\end{equation} 
where $\beta:=\frac{\alpha\max\{u_{\max}^{2},u_{\min}^{2}\}}{2(1-\alpha)}$ is an adjustable weight to penalize
travel time relative to the energy cost of CAV $i$. 

A central Roadside unit (RSU) receives the state and control information $[x_i(t), v_i(t), u_i(t)]^T$ from CAVs through vehicle-to-infrastructure (V2X) communication and stores them in a table as shown in Fig. \ref{fig:intersection}. It is assumed that the coordinator knows the entry and exit lanes for each CAV upon their arrival and uses them to determine
the list of MPs from the set $\{M_1, \dots, M_{24}\}$ (shown in Fig. \ref{fig:intersection}) in its planned trajectory. It facilitates safe coordination by providing each CAV with relevant information about other CAVs in the network, that the CAV has to yield to while traveling through the CZ. It does so by assigning each CAV a unique index based on a passing sequence policy and, tabulates and stores the information of the CAVs according to the assigned indices as shown in Fig. \ref{fig:intersection}. Let $S(t)$ be the set of CAV indices in the coordinator queue table and $N(t) = |S(t)|$ be the total number of CAVs in the CZ at time $t$. 
The default passing sequence is implemented using First In First Out (FIFO) policy which assigns $N(t)+1$ to a newly arrived CAV, and decrements the indices of all CAV with index greater than $i$ by 1, when CAV $i$ exits the CZ.


\subsection{\textbf{Constraints/Rules in the Control Zone}}
\label{rules}

The following section summarizes the rules that CAVs in the CZ must follow to navigate safely through the intersection.

\noindent{\bf Constraint 1} (Rear-End Safety Constraint): Let $i_{p}$ denote the index of the CAV which physically immediately precedes CAV $i$ in the CZ (if one is present). It is required that CAV $i$ conforms to the following constraint:
\begin{equation}
 x_{i_p}(t) - x_i(t) - \varphi v_{i}(t) - \Delta\geq 0,\text{ \ }\forall t\in[ t_{i}^{0},t_{i}^{f}] \label{Safety}%
\end{equation}
where $\varphi$ denotes the reaction time and $\Delta$ is a given minimum safe distance which depends on the length of these two CAVs.

\noindent{\bf Constraint 2} (Safe Merging Constraint): Every CAV $i$ should leave enough room for the CAV preceding it upon arriving at a MP, to avoid a lateral collision i.e.,
\begin{equation}
\label{SafeMerging}
x_{i_m}(t_{i}^{m}) - x_{i}(t_{i}^{m}) - \varphi v_{i}(t_{i}^{m}) - \Delta\geq 0,
\end{equation}
where $i_m$ is the index of the CAV that may collide with CAV $i$ at the merging points $m \in \mathcal{M}_i$ where $\mathcal{M}_i \subset \lbrace M_1,...,M_{24} \rbrace$, $\mathcal{M}_i$ is the set of MPs that CAV $i$ passes in the CZ, and $t_i^m$ is time of arrival of CAV $i$ at the MP. 

\noindent{\bf Constraint 3} (Vehicle Limitations): Finally, there are constraints on the speed and acceleration for each $i\in S(t)$:
\begin{equation}
\label{VehicleConstraints1}%
\begin{aligned} v_{\min} \leq v_i(t)\leq v_{\max}, \forall t\in[t_i^0,t_i^f]\end{aligned}
\end{equation}
\begin{equation}
\label{VehicleConstraints2}%
\begin{aligned} u_{{min}}\leq u_i(t)\leq u_{{max}}, \forall t\in[t_i^0,t_i^f] \end{aligned}
\end{equation}
where $v_{min} \geq 0$, $v_{max} > 0$ denote the minimum and maximum speed, and $u_{min} < 0$ and $u_{max}>0$ denote the minimum and maximum control respectively.


\section{Threat model}
\label{threat_model}
The adversarial effects of malicious attacks have been highlighted in our preliminary study in \cite{Ahmad_01}, namely, creating traffic jams across multiple roads due to the cooperative aspect of the control scheme, and in the worst case accidents, thus warrant making the control robust against these attacks.

\begin{definition}
    (Safe coordination) In our context, it is defined as the ability of the coordination and control framework to guarantee the satisfaction of \eqref{Safety} and \eqref{SafeMerging} for every CAV $i \in S(t) \ \forall t$ by conforming to \eqref{VehicleConstraints1} and \eqref{VehicleConstraints2}, to navigate through the CZ without any collision.
\end{definition}

\begin{definition}
    (Uncooperative vehicle) We define a CAV $i \in S(t)$ as \emph{uncooperative} if its free-flow speed is abnormally low in the CZ i.e. $v_i(t) \leq v_{low}$ (where $v_{low}$ is considered abnormally low for the CZ), thus worsening traffic throughput.
\end{definition}

\begin{definition}
    (Adversarial agent) An agent is called adversarial if it has one of the following objectives: (i) prevent \textit{safe coordination}, (ii) \emph{reduce traffic throughput}, by introducing \emph{cyber-attacks}. 
\end{definition}

Note that adversarial agents introduce attacks with malicious intent, whereas uncooperative CAVs are not malicious and may be going slow due to various reasons like faults, failures, and so on. 

\begin{assumption}
\label{accident_assumption}
    Adversarial agents do not collide with other CAVs, nor do they attempt to cause collisions between CAVs and themselves.
\end{assumption}

\subsection{Sybil Attack:}

A single malicious client (could be a CAV or attacker nearby the $\mathrm{CZ}$) may spoof one or multiple unique identities and register them in the coordinator queue table. We assume at any time $t$, there are two groups of CAVs in CZ: i. Normal CAVs and ii. fake CAVs. Let $S_x(t)$ and $S_s(t)$ be the set of the indices of normal and fake CAVs in the FIFO queue of the coordinator unit. Therefore at any time $t$, there are $N(t)=\left|S_x(t)\right|+\left|S_{s}(t)\right|$ CAVs which communicate their state and control information to the RSU. There can be one or more fake clients/CAVs in the $\mathrm{CZ}$ at any time $t$.

A Sybil attack is one where the $S_{s} (t) \subset S(t)$ is a nonempty set that is located in the coordinator queue table, but unknown to the coordinator.
For example, Fig. \ref{fig:intersection} presents a scenario, where there are multiple fake CAVs with indices $S_{s}(t) = \{3,5\}.$

\begin{assumption}
\label{max_count_assumption}
    There is a limit on the number of fake CAVs that an adversary can spoof during a Sybil attack due to resource and energy limitations.
\end{assumption}

\section{Trust framework}
\label{trust_framework}
In this section, we present our trust framework inspired from the ideas in \cite{Cheng_01,Cheng_02,Hu_01}. We consider that the central coordinator is trustworthy, and monitors, computes and stores the trust of every CAV $i \in S(t)$ in the network at every time $t$ denoted as $\tau_i(t) \in [0,1]$. The trust is determined based on identified behavioral specifications specific to the CAVs in the CZ, which are described below.

\noindent\textbf{Behavioral Specifications:}
 
\begin{enumerate}
    \item \label{vision_perception} \textbf{Co-observation consistency checks}:  
    Based on the reported position of the CAVs, for each CAV $i$ the coordinator identifies a set $S_{i}^{o}(t)$ of CAVs that CAV $i$ should be visible to at time $t$. Let $S_{i}^{\hat{o}}(t)$ be the set of CAVs which report estimated states of CAV $i$. Then the specification is $S_{i}^{o}(t) = S_{i}^{\hat{o}}(t)$. 
    
    \item \textbf{Initial condition checks}: The reported initial states particularly the position information of the CAVs has to be consistent. 
    
    \item \textbf{Dynamic model checks}: 
    The physical model similar to \eqref{VehicleDynamics} is invariant 
    and hence, the data communicated by each CAV has to always satisfy the underlying model. 
    \item \textbf{Control zone rule checks:} The rules for safe coordination and the vehicle limitations presented in \ref{rules} are invariant and mandatory for every CAV in the CZ. Hence, the specification is, every CAV $i \in S(t) \ \forall t$ has to conform to all rules in \ref{rules} while in the CZ. 
\end{enumerate}

Let $\mathcal{B}$ be the index set of the behavioral specifications in the order they are enumerated above. For each CAV $i \in S(t), \ \forall t \in [t_i^0,t_i^f]$ the coordinator assigns positive evidence $r_{i,j}(t)$ and negative evidence $p_{i,j}(t)$ for conformance and violation of every specification $j \in \mathcal{B}$ respectively (where $0 \leq r_{i,j}(t) \leq r_{max}, 0 \leq p_{i,j}(t) \leq p_{max}$), which it uses to update $\tau_i(t)$. We define $R_i(t)$ and $P_i(t)$ as cumulative positive and negative evidence for CAV $i$ at time $t$ discounted by trust of other CAVs (if the check involves another CAV, like in \eqref{Safety} and \eqref{SafeMerging}, as they can be untrustworthy). We also define a time discount factor $\gamma \in (0,1)$ as defined in \eqref{evidence}. In addition, we have a non-informative prior weight $h_i$ as in \cite{Cheng_01,Cheng_02}. Let the set of checks for every CAV involving another CAV(s) be denoted as $\mathcal{B}_{a} \subset \mathcal{B}$. The set of other CAVs involved in check $j \in \mathcal{B}_a$ when applied to CAV $i$, is denoted as $S_{i,j}(t) \subseteq S(t)/\{i\}$. 
Then, the trust metric 
is updated as follows: 

\begin{equation} \label{trust}
    \tau_i(t) = \frac{R_i(t)}{R_i(t) + P_i(t) + h_i} \ \ \forall i \in S(t)
\end{equation}

\begin{align} \label{evidence}
    R_i(t) = & \gamma R_i(t-1) + \sum_{j \in \mathcal{B} \backslash \mathcal{B}_{a}} r_{i,j}(t) + \sum_{j \in \mathcal{B}_{a}}\prod_{k\in S_{i,j}} \tau_k(t)r_{i,j}(t)  \nonumber \\
    P_i(t) = &\gamma P_i(t-1) + \underbrace{\sum_{j \in \mathcal{B} \backslash \mathcal{B}_{a}} p_{i,j}(t) + \sum_{j \in \mathcal{B}_{a}} \prod_{k\in S_{i,j}} \tau_k(t)p_{i,j}(t)}_{p_i(t)} \nonumber \\
    &\forall i \in S(t), \forall t \in [t_i^0,t_i^f]
\end{align}

Finally, we define a lower trust threshold $\delta \in (0, 1/2)$, and a higher trust threshold $1 - \delta$ for subsequent sections. It is important to emphasize that, in practice, the magnitude of negative evidence is different and significantly higher compared to the magnitude of positive evidence. This model of trust relationships considers the social aspect, where a single action can cause significant damage to a trust relationship, and recovery from such damage is challenging \cite{Garlichs_01}. 

\begin{remark}
    Our implementation is agnostic to the specific implementation of the trust framework and the ideas can be used for any framework provided that the trust metric can accurately encapsulate the behavioral specification of the network and distinguish between normal and anomalous behavior for every CAV in real-time.
\end{remark}

\section{Safe and Resilient Control Formulation}
\label{resilient_control_formulation}
We adopt a decentralized \emph{Optimal Control Problem} (OCP) controller for the CAVs that uses Control Barrier Functions (CBF). CBFs provide manifold benefits namely, i. their forward invariance property guarantees satisfaction of the constraints of the OCP, and ii. they transform the original constraints to linear constraints in terms of the control input which makes them computationally efficient, thus, attractive for real-time applications \cite{Xiao_03}.

\textbf{The OCBF Controller} \cite{Xiao_03}. Firstly, Control Barrier Functions (CBFs) that ensure the constraints (\ref{Safety}), (\ref{SafeMerging}), \eqref{VehicleConstraints1} and \eqref{VehicleConstraints2} are derived, subject to the vehicle dynamics in (\ref{VehicleDynamics}) by defining $f(\boldsymbol{x}_i(t))=[v_i(t),0]^T$ and $g(\boldsymbol{x}_i(t))=[0,1]^T$. Each of these constraints can be easily written in the form of $b_q(\boldsymbol{x}(t)) \geq 0$, $q \in \lbrace 1,...,n \rbrace$ where $n$ stands for the number of constraints only dependent on state variables and $\boldsymbol{x}(t)=[\boldsymbol{x}_1(t),\boldsymbol{x}_2(t),...,\boldsymbol{x}_{N(t)}(t)]$. The CBF method (details provided in \cite{Xiao_03, Xiao2019}) maps a constraint $b_q(\boldsymbol{x}(t)) \geq 0$ onto a new constraint which is \emph{linear} in the control input and takes the general form
\begin{equation} \label{cbf_condition}
L_fb_q(\boldsymbol{x}(t))+L_gb_q(\boldsymbol{x}(t))u_i(t)+\kappa_q( b_q(\boldsymbol{x}(t))) \geq 0.
\end{equation}
where. $\kappa_q$ is a class $\mathcal{K}$ function.

A Control Lyapunov Function (CLF) is used for velocity tracking with $v_{i}^{ref}(t)$ as the reference by setting $V(\boldsymbol{x}_i(t))=(v_i(t)-v_{i}^{ref}(t))^2$, rendering the following CLF constraint:
\begin{equation}\label{CLF_constraint}
L_fV(\boldsymbol{x}_i(t))+L_gV(\boldsymbol{x}_i(t))\boldsymbol{u}_i(t)+ c_3 V(\boldsymbol{x}_i(t))\leq e_i(t),
\end{equation}
where $e_i(t)$ makes this a soft constraint. \emph{Note that} the CBFs are used to enforce hard constraints mentioned in section \ref{rules}, whereas CLFs are used to enforce soft constraints. 

The OCBF problem corresponding to \eqref{eqn:energyobja} is formulated as:
\begin{equation}\label{QP-OCBF}\small
\min_{u_i(t),e_i(t)}J_i(u_i(t),e_i(t)):=\int_{t_i^0}^{t_i^f}\big[\frac{1}{2}(u_i(t)-u_{i}^{ref}(t))^2+\lambda e^2_i(t)\big]dt
\end{equation}
subject to vehicle dynamics (\ref{VehicleDynamics}), the CBF constraints \eqref{cbf_condition}, $\forall q=\{1,...,n\}$ and CLF constraint \eqref{CLF_constraint}. In this approach,(i) $u_i^{ref}$ is generated by solving the unconstrained optimal control problem in \eqref{eqn:energyobja}, (ii) the resulting $u_i^{ref}$ is optimally tracked such that constraints including the CBF constraints \eqref{cbf_condition} $\forall q=\{1,...,n\}$ are satisfied. We can solve this dynamic optimization problem by discretizing $[t_i^0,t_i^f]$ into intervals $[t_i^0,t_i^0+t_s],...,[t_i^0+kt_s,t_i^0+(k+1)t_s],...$ with equal length $t_s$ and solving (\ref{QP-OCBF}) over each time interval through solving a QP at each time step:
\begin{align} \label{QP}
\min_{u_{i,k},e_{i,k}}&[ \frac{1}{2}(u_{i,k}-u_i^{ref}(t_{i,k}))^2+\lambda e_{i,k}^{2}]
\end{align}
subject to the CBF constraints \eqref{cbf_condition}, $\forall q=\{1,...,n\}$, CLF constraint (\ref{CLF_constraint}) and dynamics \eqref{VehicleDynamics}, where all constraints are linear in the decision variables.


\subsection{Resilient Control and Coordination Scheme}

We propose a resilient coordination and control scheme to mitigate the adversarial effects in terms of causing (i) collision and (ii) traffic congestion. Resilience is the ability of the framework to guarantee \textit{safe coordination} and \textit{mitigate any traffic jam} introduced by adversarial agents and uncooperative CAVs.

\subsubsection{\noindent \textbf{Resilience goal (collision avoidance)}}

\

\noindent {\textbf{Trust-based search:}}  The coordinator incorporates trust besides the default passing sequence policy (e.g. FIFO) to identify indices of CAVs, any CAV may conflict within the CZ based on \eqref{Safety} and \eqref{SafeMerging}. Under the default passing sequence, for every CAV $i \in S(t)$, 
the coordinator has to identify indices of all CAVs which includes i. index of the CAV that immediately precedes CAV $i$ physically in its lane and ii. index of the CAV that will precede $i$ immediately at every $m \in \mathcal{M}_i$ in the intersection. For example, in Fig. \ref{fig:intersection}, $\mathcal{M}_6 = \{M_{11}, M_{13}, M_{16}, M_{20}\}$, and as per FIFO sequencing, $6_{M_{20}} = 5$, since CAV 5 is the CAV that will precede it. 

The trust-based search process identifies all the CAVs that will precede $i$ until the first CAV whose trust value is greater than or equal to $1- \delta$ and forms a set $S_{i,m}(t) \subset S(t)$ containing all the CAV indices identified during the search process. It follows the same search process for every MP in $\mathcal{M}_i$ and also for \eqref{Safety}. Therefore, for each CAV $i$, the coordinator identifies $S_{i}^p(t) \subset S(t)$, and $S_{i}^{M}(t) = \bigcup_{m \in \mathcal{M}_i} S_{i,m}(t)$ (where $S_{i}^p(t)$ is the set for \eqref{Safety} and $S_{i}^{M}(t)$ correspond to the set of indices for every MP). The search process is formalized as follows:
\begin{align}
    &S_m(t) = \{i_+ \in S(t) |\ i_+ < i, m \in \mathcal{M}_i\}  \\
    &k_{min} = \text{min} \ \{k \in S_m(t)| \tau_k \geq 1- \delta\}  \label{searchindex}\\
    &\tilde{S}_{i,m}(t) = S_m(t,1) \ \ \ \ \ \ \ \label{default search} \\
    &S_{i,m}(t) = \cup_{k = 1}^{k_{min}} S_{m}(t,k)  \label{trust-based search} 
\end{align}

\noindent where $S_{i,m} (t,k)$ is the $k-th$ element of set $S_{i,m}(t)$. The set returned by the default search process is given in \eqref{default search}, and the trust based search returns the set in \eqref{trust-based search}. Note that, there are three scenarios possible from the search process: i. $k_{min} = \emptyset$ meaning there are no constraints for MP $m$, ii. $\tilde{S}_{i,m} = S_{i,m}$ when $k_{min} = S_m(t,1)$ meaning that the trust of the CAV immediately proceeding CAV $i$ at $m$ is greater than or equal to $1-\delta$, and iii. $k_{min} > S_m(t,1)$, hence $\tilde{S}_{i,m}(t) \subset S_{i,m}$ implying that the immediately preceding CAV has trust lower than $1-\delta$. For the example in Fig. \ref{fig:intersection}, notice $4_p = 3$. However, since $\tau_3 < 1-\delta$, the search process will continue and return $S_{4,p} = \{3,1\}$. Similarly, under the trust-based search scheme $6_{M_{20}} = \{5,4,3,2\}$ as CAVs 2, 3 4, and 5 have trust less than $1- \delta$.


The state and control information of the CAVs in $S_i^p(t) \cup S_{i}^{M}(t)$ are communicated to CAV $i$ at each $t$, and the corresponding CBF constraints for each CAVs are incorporated to the control in \eqref{QP}.  


\begin{lemma}
\label{feasibility_lemma}
    The introduction of additional constraints due to \emph{trust-based search}, (including those due to default search process) in the control for any CAV $i \in S(t)$ in \eqref{QP} at time $t'$ where $t' \in [t_{i}^{0},t_{i}^{f}]$ does not affect the feasibility of the problem \eqref{QP} at $t'.$ 
\end{lemma}
\ifITSC
\begin{proof}
    As mentioned, for any CAV $i \in S(t)$, $S_{i,m}(t) \subset \tilde{S}_{i,m}(t)$ is the set of indices of the CAVs that $i$ needs to stay safe to at MP $m \in \mathcal{M}_i$ under trust based search scheme. Let the trust-based search adds an index of a CAV $i_- (< i)$ to $S_{i,m}(t)$. We define, $i_1 = S_{i,m}(t,1)$, $b_{i,i_1}(\boldsymbol{x}(t^{'})) = x_{i_1}(t') - x_{i}(t') - \varphi v_{i}(t') - \Delta$. Similarly, $b_{i_1,i_-}(\boldsymbol{x}(t^{'}))$ and $b_{i,i_-}(\boldsymbol{x}(t^{'}))$ can be defined.

    Notice that, $m \in \mathcal{M}_{i_-}$. Also notice, $i_- < i_1 < i$, since $i_-$ will cross MP $\mathcal{M}$ before $i_1$ which will cross before $i$. This implies, $b_{i_1,i_-}(\boldsymbol{x}(t^{'})) \geq 0 \text{ and }, b_{i,i_1}(\boldsymbol{x}(t^{'})) \geq 0$. Hence $b_{i,i_-}(\boldsymbol{x}(t^{'}))  = b_{i_1,i_-}(\boldsymbol{x}(t^{'})) + b_{i,i_1}(\boldsymbol{x}(t^{'})) \geq 0$, implying the constraint is initially feasible and $i$ is safe to $i^-$ at $t'$. Hence, the addition of a new CBF constraint due to \emph{trust-based search} corresponding to $i_-$ to the control of CAV $i$ (or any CAV) in \eqref{QP} doesn't affect the feasibility at time $t'$.
\end{proof}
\fi

\begin{remark}
    Lemma 1 is necessary for guaranteeing the satisfaction of the CBF constraints corresponding to the CAVs returned by \emph{trust-based search} process $\forall \ t\geqslant t'$ using the forward invariant property of CBFs \cite{Xiao_03}.
\end{remark}

\begin{theorem} Given $0 \leq r_{i,j}(t) \leq r_{max} \ \forall t, \ \forall i \in S(t), \ \forall j \in \mathcal{B},$ the introduction of trust-based search guarantees avoidance of collision by guaranteeing the satisfaction of \eqref{Safety} and \eqref{SafeMerging} that can be caused by adversarial agents.
\end{theorem}
 
\ifITSC
\begin{proof} Let, adversarial CAV $i \in S(t) /\{k\}$ attempts to induce an accident to CAV $k$ in the CZ at time $t$ through using one of the attacks in section \ref{threat_model}. Firstly, notice $k$ must be greater than $i$, else it is impossible to create an accident due to i. each CAV staying safe to all immediately preceding CAVs in their trajectory, and ii. assumption \eqref{accident_assumption}. At some time $t>t_i^0$, CAV $i$ has to violate its own constraint; else, if $i$ satisfies its own constraint, so will each CAV $i_+ \in S(t)$ ($i_+ = \{i_+ \in S(t) |\ i_+ >i, (\mathcal{M}_i \cap \mathcal{M}_{i_+}) \neq \emptyset\}$) queuing behind $i$ and so does CAV $k$.  Upon violation of a constraint by CAV $i$, two scenarios can occur. 

Case (i) $\tau_i(t) > 1-\delta$:  Given $r_{i, j}(t) < r_{max}$,
\begin{align}
&\sum_{j \in \mathcal{B} \backslash \mathcal{B}_{a}} r_{i,j}(t) + \sum_{j \in \mathcal{B}_{a}} \prod_{k\in S_{i,j}^a} \tau_k(t)r_{i,j}(t) \leq \left|\mathcal{B}\right| r_{max} \nonumber\\
& \therefore R_i(t) \leq\left|\mathcal{B}\right| r_{max}+\gamma R_i(t-1) \leq \frac{\left|\mathcal{B}\right| r_{max}}{1-\gamma} \nonumber\\
& \text{and}, p_{i, j}(t) \geqslant 0 \Rightarrow P_i(t) \geqslant 0 \nonumber
\end{align}
We need the trust $\tau_i < 1 - \delta$ immediately to trigger trust-based search. Hence we need to show given $\tau_i(t) > 1-\delta$, $\exists p_{i}(t+1)$ and $p_{i,j}(t+1)$ s.t. $\tau_i(t+1)<1-\delta$ i.e. $trust-based \ search$ is triggered in next iteration.

\begin{align}
& \tau_i(t+1)=\frac{R_i(t+1)}{R_i(t+1)+P_i(t+1)+h_i} < 1-\delta \nonumber\\
& \Rightarrow P_i(t+1)>\frac{R_i(t+1)}{1-\delta}-R_i(t+1) - h_i \nonumber\\
& \Rightarrow p_i(t+1)+\gamma P_i(t)>\frac{\delta}{1-\delta} \frac{|\mathcal{B}|r_{max}}{1-\gamma} \nonumber\\
& \Rightarrow p_i(t+1)>\frac{\delta}{1-\delta} \frac{|\mathcal{B}|r_{max}}{1-\gamma}\geq\frac{\delta}{1-\delta} \frac{|\mathcal{B}|r_{max}}{1-\gamma}-\gamma P_i(t) \nonumber
\end{align}

\noindent and, when $\tau_i(t+1) < 1-\delta$,  then, CAV $k$ will stay safe from $i$, and, as well to all other CAVs that will arrive before $i$ as well as all CAVs in $S_i^p(t) \cup S_i^M(t)$. We set the sampling time to be in the order of $ms$, combining this with Lemma \ref{feasibility_lemma} will guarantee safety for CAV $k$, thus preventing any collision.

Case (ii) $ \tau_i(t)<1- \delta$: The same argument in the preceding paragraph apply, and hence guarantee safety for CAV $k$. A similar argument can be extended to guarantee safety for every CAV $i_+ \in S(t)$ which completes the proof.
\end{proof}

\fi

\ifITSC
\subsubsection{\noindent \textbf{Resilience goal (traffic jam avoidance)}} The goal is to avoid traffic buildup in the network due to uncooperative/malicious agents acting deliberately to create traffic congestion in the network.  

\textbf{Robust Scheduling:}
We propose a central, $event-driven$, $robust \ scheduling$ scheme that implements FIFO passing sequence for the CAVs in the CZ during normal operation; however, reschedule the CAVs in the presence of adversarial CAVs to prevent any traffic jam. We define a rescheduling zone in the CZ of length $L_1$ as shown in Fig. \ref{fig:intersection}. 
We first present the rescheduling schemes followed by the events resulting in CAV scheduling (rescheduling).

\noindent{\textbf{Trust-aware scheduling:}} 
Under this scheme, CAVs are indexed (sequenced) in descending order of their trust value, which is intended to encourage CAVs to act in a manner that earns them trust as quickly as possible upon arrival in the CZ. The algorithm is presented in Algorithm \ref{Trust_rescheduling_algorithm}.
\begin{algorithm} [b]
\SetKwInOut{Input}{Input}\SetKwInOut{Output}{Output}
\SetKwComment{comment}{\#}{}
\caption{Trust-aware rescheduling algorithm}
\label{Trust_rescheduling_algorithm}
\Input{$\tau_i(t), \tau_i(t-1) \ \forall i \in S(t),$  \ 
$\mathcal{A}$ = \text{allowable proportion of CAVs with low trust} 
}
\Output {New sequence}
\BlankLine
Set of CAVs with low trust $S_R(t) = \emptyset$ \\
\For {each CAV $i$ in Rescheduling zone}
{
    \If{${\tau}_i\left(t-1\right)- {\tau}_i\left({t}\right) \geq 0 \ \& \ {\tau}_i\left(t\right) \leq \delta$}
        {append $i$ to $S_R\left(t\right)$
        }
}

\If{$\lvert S_R(t)\rvert \geq \mathcal{A} \times N(t)$} 
{
    Solve \eqref{trust_based_resequencing}
}
\end{algorithm}

The problem of the rescheduling (i.e. finding a passing sequence) based on the trust score of the CAVs is formulated as a Integer Linear Program (ILP) as in \eqref{trust_based_resequencing}. We define the index of the first CAV in the queue to re-sequence from as $k_{min} = \min S_R(t)$ (where $S_R(t)$ is defined in Algorithm \ref{Trust_rescheduling_algorithm}) and $S_+(k_{min}) = \{k_{min},\dots,N(t)\}$, as the set of indices of the CAVs to be rescheduled in $S(t)$. 

\begin{align}
    \argmax_{\left\{a_i \in S_+(k_{min})\right\}} &\sum_{i=k_{min}}^{N(t)} (1-{\tau}_i\left(t\right))a_i     \label{trust_based_resequencing} \\
    \textnormal{s.t.} \ \ &  a_j - a_k \geq \nu \ \ \forall j \in S_+(k_{min}), k \in S_{j}^p \label{rearend_ILP} \\
    &a_j \neq a_k \ \ j,k \in S_+(k_{min})  \label{unique_ILP}\\
    & \nu \geq 1  \label{feasibleset_ILP} 
\end{align}

where \eqref{rearend_ILP} corresponds to constraint \eqref{Safety}, $\{a_{k_{min}},\dots, a_{N(t)}\}$ are the new indices of the CAVs in $S_+(k_{min})$. For example in Fig. \ref{fig:intersection}, rescheduling moves CAV 3 (and immediately preceding CAV 4) down in the queue beneath the remaining CAVs in the CZ since $\tau_3$ is the lowest of all CAVs in the Rescheduling zone.  

\noindent\textbf{Lane-priority based rescheduling:} This idea is based on lane priority assignment where lanes are prioritized by observing the number of uncooperative CAVs in that lane. However, note that the presence of slow CAVs in the trajectory of a particular CAV $i$ (i.e. constraints of CAV $i$) can also cause it to go slower than $v_{low}$. Hence, we identify any CAV $i \in S(t)$ as uncooperative at time $t$, if $v_i(t) \leq v_{low}$ and $\nexists i_+ \in S_{i}^{M}$ s.t. $v_{i_+}(t) \leq v_{low}$; we group the slow moving CAVs at time $t$ for lane $l$ into the set $S_{l}^a(t)$ where $l \in \{l_1, \dots, l_8\}$. Following that we compute \emph{the priority of any lane} $l$ using the following equation.

\begin{equation}
\label{lane_priority}
    \zeta_l(t) = 1- \frac{S_{l}^a(t)}{\sum_{l \in [l_1,\dots,l_8]}S_{l}^a(t)+c}, \ \ c (\approx 0) \in \mathbb{R}^+
\end{equation}

We define $k_{min} = \min \{k | k \in S^a(t)\}$ and $S_+(k_{min}) = \{k_{min},\dots,N(t)\}$, where $S^a(t) = \cup_{l \in \{l_1, \dots, l_8\}}S_{l}^a(t)$. We also define a set $S_+^r(t)$ containing the indices of CAVs that are not physically following any slow moving CAV:

\begin{equation}
    S_+^r(t) = \{i \in S_+(k_{min})|i_p(t) \cap S^a(t) = \emptyset\} \nonumber
\end{equation}

Then, we define the following condition that triggers the re-sequencing event:

\begin{equation}
\label{lane_priority_resequencing_event}
    \frac{|S_+^r(t)|}{|S_+(k_{min})|} \geq \mathcal{A}_l, \ \ \mathcal{A}_l \in \mathbb{R}^+ \text{is a preset threshold}
\end{equation}

The re-sequencing is done by solving the following ILP that returns the new indices of the CAVs in $S_+(k_{min})$

\begin{align}
    &\argmax_{\left\{a_i \in S_+(k_{min})\right\}} \sum_{i \in S_a} (1-{\zeta_{i}^l}\left(t\right))a_i \nonumber \\
    & \ \ \ \ \ \ \ \ \ \ \  \ \text{s. t.} \eqref{rearend_ILP}, \eqref{unique_ILP} \text{ and } \eqref{feasibleset_ILP}.
    \label{lane_priority_resequencing}
\end{align}

where $\zeta_{i}^l(t)$ is the priority associated to the lane that CAV $i$ is physically located at time $t$ which can be found in \eqref{lane_priority}, and $\{a_{k_{min}},\dots, a_{N(t)}\}$ are the new indices of the CAVs in $S_+(k_{min})$. For example, in Fig. \ref{fig:intersection}, velocities of CAV 2 and 5 are $v_2 < v_{low}$ and $v_5 < v_{low}$ in lane $l_8$ and $l_5$ respectively. Hence, $S^{a}_{l_8} = \{2, 8, 9\}$ and $S^{a}_{l_5} = \{5, 6\}$. Therefore, the priorities of $l_8$ and $l_5$ become $0.4$ and $0.6$ respectively, while all other lane priority remains equal and high. This causes CAVs in $S^{a}_{l_8}$ to be moved to the very end of the queue followed by $S^{a}_{l_5}$, as seen in the table in Fig. \ref{fig:intersection}.

\begin{lemma}
    The rescheduled sequence is guaranteed to be feasible if $ L - L_1 \geq \frac{v_{\max}^2}{2|u_{\min}|} + \Delta$, where $\Delta$ is defined as in \eqref{Safety} and \eqref{SafeMerging}.
\end{lemma}

\ifITSC
\begin{proof}The maximum velocity for any CAV $i$ is $v_{\max}$, and the maximum deceleration is $|{u_{\min}}|$. Thus the minimum distance required to come to full stop for any CAV $i$ is $\frac{v_{\max}^2}{2|u_{\min}|}$. Hence, to satisfy the constraint \eqref{Safety} and \eqref{SafeMerging}, the minimum distance between the merging point and the end of the re-sequencing zone has to be greater than or equal to $\frac{v_{\max}^2}{2|u_{\min}|} + \Delta$, which will guarantee the feasibility of rescheduled sequence.
\end{proof}
\fi

The list of $events$ that cause scheduling (rescheduling) CAVs in the CZ are enumerated below:

\begin{enumerate}
    \item \textbf{Arrival event:} This corresponds to a CAV that has just arrived at the CZ, and hence has to be added in the queue Table \ref{fig:intersection} and an index has to be assigned to it using the default sequencing scheme (FIFO).
    \item \textbf{Departure event:} An exiting CAV triggers this event, after which the row corresponding to that CAV is removed from the coordinator table and the indices of all CAVs are decreased by 1.
    \item \textbf{Reschedule event:} This corresponds to an event,  
    when the presence of uncooperative CAVs triggers the event as a result of the condition in \eqref{lane_priority_resequencing_event} , or, the presence of low trustworthy CAVs results in trust-based rescheduling as in algorithm \ref{Trust_rescheduling_algorithm}. 
\end{enumerate}

\begin{remark}
    Notice, the two robust rescheduling schemes are event-driven, and hence, can be simultaneously incorporated. 
\end{remark}

The default scheduling scheme in \cite{Xu_02} and the presented rescheduling schemes render a unique index list for the CAVs in the CZ based on which they cross the intersection in descending order of their indices. 

\fi

\section{Simulation Results}
\label{Results}
In this section, we present the results of our proposed resilient control and coordination scheme 
for the threats mentioned in section \ref{threat_model}. We performed the simulations in Matlab and \textsc{ode45} to integrate the CAV dynamics. The value of $\delta$ was set to 0.1. The positive and negative evidence magnitudes for the tests in the order they are mentioned in section \ref{trust_framework} are: $r_i(t) = [0.6,0.6,0.6,0.6]^T$ and $p_i(t) = [1000,100,50,1]^T \ \forall i \in S(t)$ and $\forall t$. The intersection dimensions are: $L= 300\textnormal{m}$, $A = 30\textnormal{m}^2$; and, the remaining parameters are $\varphi = 1.8\textnormal{s}$,  $\Delta = 3.78 \textnormal{m}$, $\beta_1= 1, u_{\max} = 4.905 \textnormal{m/s}^2, u_{\min} = -5.886\textnormal{m/s}^2, v_{\max} = 108 \textnormal{km/h}, v_{\min} = 0 \textnormal{km/h}$. 
Finally, we also adopted a realistic energy consumption model from \cite{Xu_02} to supplement the simple surrogate $L_2$-norm ($u^2$) model in our analysis.


\subsection{Resilient Control and Coordination}

\begin{figure}[h]
\begin{center}
 \includegraphics[scale = 0.45]{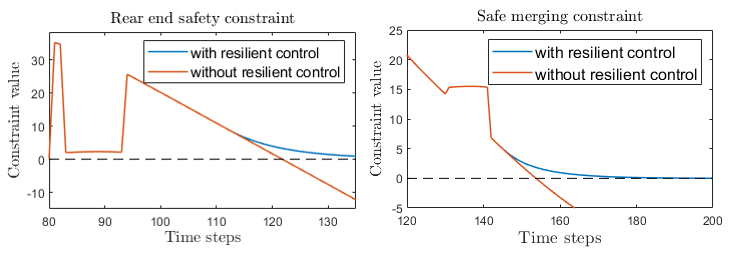}
\caption{Comparison of rear-end and lateral constraint value given in \eqref{Safety} and \eqref{SafeMerging},  for a real CAV with respect to another real CAV, with and without the proposed resilient control scheme.
} \label{fig:resilient_control_2}%
\end{center}
\end{figure}

\textbf{Resilience control}: Presented in Fig. \ref{fig:resilient_control_2} are results for the scenarios when a fake CAV attempts to violate safety constraints between real CAVs with the aim of creating an accident. The plot shows the value of the safety constraints that the fake CAV attempts to violate with and without our proposed resilient control scheme for \emph{safe coordination}. As can be seen, both rear collision and collision at merging point inside the intersection (which can cause traffic disruption and jam inside the intersection) is possible which is eliminated through our proposed safe and resilient control and coordination scheme, using \textit{trust-based search}. 



\ifITSC
\noindent\textbf{Lane-priority based rescheduling}: 
An extensive simulation with multiple slow CAVs was conducted to demonstrate the effectiveness and significance of our lane-priority based re-scheduling scheme with its results demonstrated in Fig. \ref{fig:lane_priority}. We introduced from 2 up to 8 uncooperative CAVs in the intersection across 3 arbitrarily chosen lanes during our simulation. As can be noticed, the cooperative nature of the algorithm can cause traffic holdups with the average travel times of CAVs from over $4$ mins. (270 secs precisely) upto around $5$ mins. However, with our proposed robust scheduling scheme based on lane priority, the average travel time was significantly reduced, and the maximum average travel time was a little over $1$ min (74 secs precisely) which was an improvement of over $3$ mins from the least average travel time without our rescheduling scheme.



\begin{figure}[h]
\begin{center}
 \includegraphics[scale = 0.4]{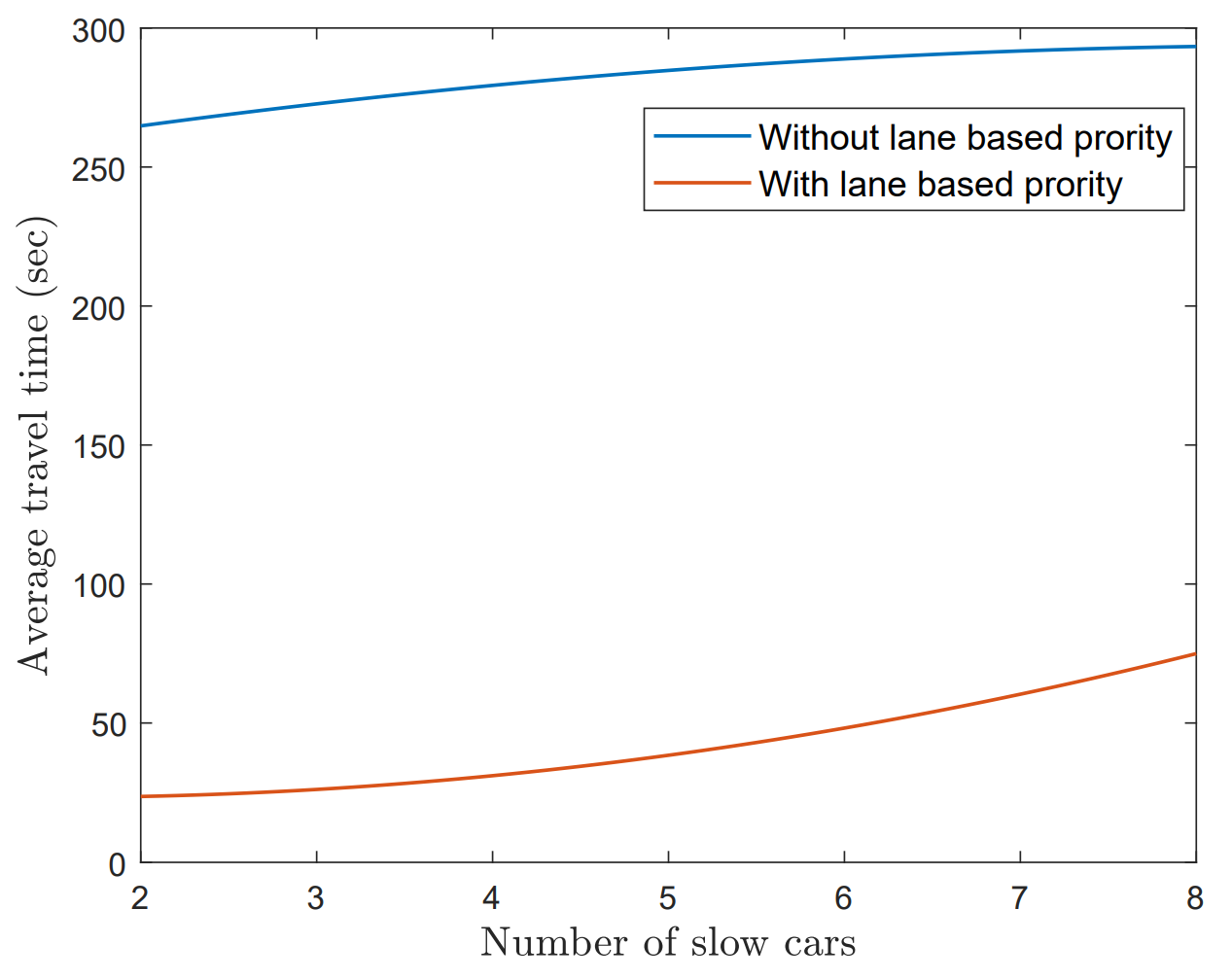}
\caption{Average time for of real CAVs for lane-based priority scheduling for various number of uncooperative CAVs.} %
\label{fig:lane_priority}%
\end{center}
\end{figure}

\noindent\textbf{Trust-aware rescheduling:} Finally, we present the results for our proposed trust-aware rescheduling scheme in Fig. \ref{fig:trust_aware_reschedule}. We introduced various percentages of fake CAVs ranging from 2\% to 15 \% fake CAVs through Sybil attack (using the model in section \ref{threat_model}). We used the various attacker models for the fake CAVs presented in \cite{Ahmad_01}. Our results demonstrate that the average travel times, energy, and fuel consumption of the real CAVs improve with the inclusion of our proposed rescheduling scheme. However, notice that the average energy eventually becomes identical, since a large proportion of spoofed CAVs cause the average travel times of the normal CAVs to increase, thus decreasing the average acceleration input (related to energy, \eqref{eqn:energyobja}). However, the average fuel consumption is improved with our proposed rescheduling scheme. Note that, eventually, as the percentage of fake CAVs approaches 100 \%, the curves for all three metrics will coincide, since all CAVs are fake.

\begin{figure*}[t]
\begin{center}
 \includegraphics[scale = 0.6]{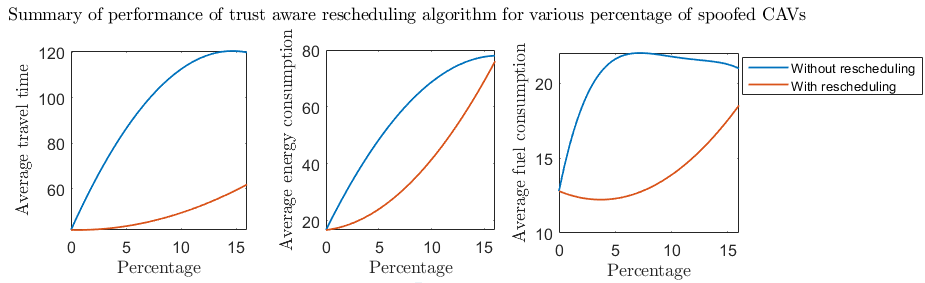}
\caption{The values of average travel time, average energy and average fuel consumption for real CAVs as a result of trust-aware rescheduling for different proportion of fake CAVs.} %
\label{fig:trust_aware_reschedule}%
\end{center}
\end{figure*}

\fi

\section{Conclusion}
\label{conclusion}
We have presented a resilient coordination and control scheme by incorporating a trust framework that offers resilience against adversarial objectives that can be introduced by malicious attacks and uncooperative CAVs. Based on our previous study we identified two main adversarial objectives namely, (i) safety violation and (ii) creating traffic congestion in the network. We used Sybil attacks to validate and demonstrate the merit of our proposed scheme which guarantees \emph{safe coordination} and can \emph{mitigate traffic jam}. In addition, we demonstrated that our proposed robust scheduling scheme, mainly, lane-priority based rescheduling can successfully mitigate the effect of uncooperative CAVs and mitigate traffic holdups introduced by them due to the cooperative coordination scheme. 
Finally, we have presented results from computer simulation to validate and demonstrate the effectiveness of our proposed attack resilient control and coordination scheme for Sybil attacks, and uncooperative CAVs.

\bibliographystyle{IEEEtran}


\end{document}